\documentclass[10pt]{article}
\usepackage{graphicx}

\def\Title#1{\begin{center} {\Large #1 } \end{center}}
\def\Author#1{\begin{center}{ \sc #1} \end{center}}
\def\Address#1{\begin{center}{ \it #1} \end{center}}

\newcommand\pubblock{\rightline{\begin{tabular}{l} Proceedings of the Fifth Annual LHCP\\ \pubnumber\\
         \pubdate  \end{tabular}}}

\newenvironment{Abstract}{\begin{quotation} \begin{center} 
             \large ABSTRACT \end{center}\bigskip 
      \begin{center}\begin{large}}{\end{large}\end{center} \end{quotation}}

\newenvironment{Presented}{\begin{quotation} \begin{center} 
             PRESENTED AT\end{center}\bigskip 
      \begin{center}\begin{large}}{\end{large}\end{center} \end{quotation}}

\def\beq{\begin{equation}}
\def\eeq#1{\label{#1}\end{equation}}
\def\eeqn{\end{equation}}

\def\beqa{\begin{eqnarray}}
\def\eeqa#1{\label{#1}\end{eqnarray}}
\def\eeqan{\end{eqnarray}}

\let\bar=\overbar

\def\Dslash{\not{\hbox{\kern-4pt $D$}}}
\def\dslash{\not{\hbox{\kern-2pt $\del$}}}

\def\msb{{\bar{\ssstyle M \kern -1pt S}}}

\usepackage{color}

\newcommand\pubnumber{CMS CR-2017/119 }

\newcommand\pubdate{\today}

\def\affiliation{
On behalf of the CMS-HCAL Collaboration, \\
Bogazici University, Bebek Istanbul, 34342, TURKEY}

\def\support{\footnote{This work was supported by Turkish Atomic Energy Authority (TAEK) and Bogazici University Research Fund Grant Number: 9620 (15B03P1). }}

\begin{document}

\large
\begin{titlepage}
\pubblock

\vfill
\Title{ CMS-HF Calorimeter Upgrade for Run II  }
\vfill

\Author{ Erhan G\"ulmez \support }
\Address{\affiliation}
\vfill
\begin{Abstract}

CMS-HF Calorimeters have been undergoing a major upgrade for the last couple of years to alleviate the problems encountered during Run I, especially in the PMT and the readout systems. In this poster, the problems caused by the old PMTs installed in the detectors and their solutions will be explained. Initially, regular PMTs with thicker windows, causing large Cherenkov radiation, were used. Instead of the light coming through the fibers from the detector, stray muons passing through the PMT itself produce Cherenkov radiation in the PMT window, resulting in erroneously large signals. Usually, large signals are the result of very high-energy particles in the calorimeter and are tagged as important. As a result, these so-called “window events” generate false triggers. Four-anode PMTs with thinner windows were selected to reduce these “window events.” Additional channels also help eliminate such remaining events through algorithms comparing the output of different PMT channels. During the EYETS 16/17 period in the LHC operations, the final components of the modifications to the readout system, namely the two-channel front-end electronics cards, are installed. Complete upgrade of the HF Calorimeter, including the preparations for the Run II will be discussed in this poster, with possible effects on the eventual data taking.
\end{Abstract}
\vfill

\begin{Presented}
The Fifth Annual Conference\\
 on Large Hadron Collider Physics \\
Shanghai Jiao Tong University, Shanghai, China\\ 
May 15-20, 2017
\end{Presented}
\vfill
\end{titlepage}
\def\thefootnote{\fnsymbol{footnote}}
\setcounter{footnote}{0}
%

\normalsize 


\section{Introduction}

The CMS-HF calorimeter improves the detection of forward jets and particles scattered in the very forward region ($3 < \eta < 5$). There are two detectors; each is placed at the end of the CMS detector. These cylindrical detectors have an active radius of 1.4 m and 1.65 m long. Plastic-clad quartz fibers are used as active elements. The long (1.65 m) fibers are optimized for the electromagnetic showers and the short ones (1.43 m) for the hadronic showers. When particles pass through the fibers, they produce Cherenkov radiation. PMTs are attached to the extended end of these fibers with the help of an air-core light-guide. 


Muons easily reach the PMTs attached to the HF detectors. These muons produce Cherenkov radiation right at the PMT windows, causing false triggers faking very high-energy events. As opposed to the regular Cherenkov light coming through the fibers, these PMT window events have larger signals and arrive a few ns earlier. At low luminosities, real events will produce comparable signals in both hadronic and electromagnetic channels, but the window events only in one channel. Two muons passing through both PMTs are very unlikely. However, in Run II luminosity has increased and so has the occupancy rate of the channels as well as the number of muons passing through the PMTs. Using the correlation of the two channels will not be enough to select the real events. 

\begin{figure}[htb]
\centering
\includegraphics[height=2in]{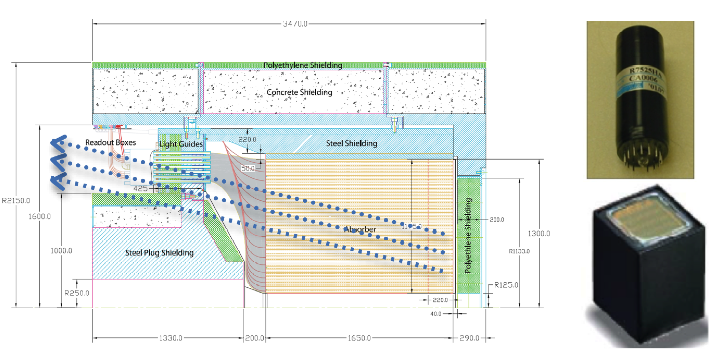}
\caption{ (left) A side drawing of the HF detector; dashed arrows represent the stray muons passing through the PMTs. (right-top) Old PMTs: R7525; Bialkali photocathode; Max. 25\% QE; Typical gain: $5{\rm x}10^5$; Window: $\sim$1.2 mm thick at the center, $\sim$6 mm thick at the edges; plano-convex geometry with face area $\sim$490~mm$^2$;  Glass jacket. (right-bottom) 4-anode PMTs: R7600U-200-M4; Ultra Bialkali photocathode; Max. 43\% QE (350 nm); Typical gain: $1.3{\rm x}10^6$; Window: $<$1 mm thick; square geometry with area $\sim$324~mm$^2$; Metal jacket. }
\label{fig1}
\end{figure}

 \section{Upgrading for Run II: Phase I upgrade}

\begin{figure}[htb]
\centering
\includegraphics[height=2in]{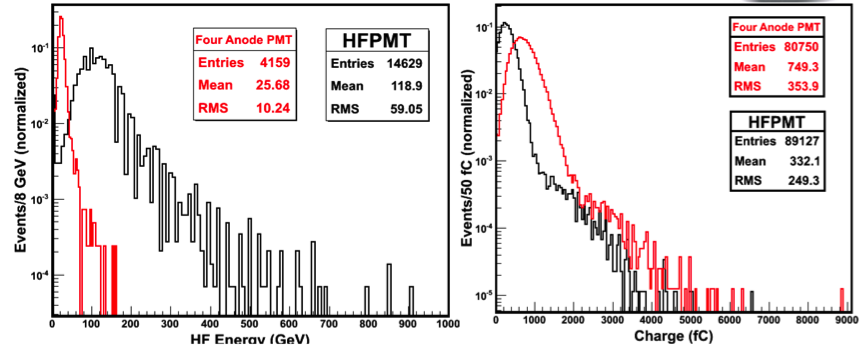}
\caption{ Raw signals from the new (red) and old PMTs (black). New PMTs produce lower window signals (left) but a higher response to the real events (right)~\cite{ref1,ref2}. }
\label{fig2}
\end{figure}

Existing PMTs (Fig.~\ref{fig1}-right-top) in the detectors are replaced with thin-window four-anode PMTs (Fig.~\ref{fig1}-right-bottom) during the first long shutdown (LS1) before Run II. Thinner windows in these PMTs produce less Cherenkov light due to the stray muons. Muons passing through the PMTs are less likely to produce signals in more than one anode as opposed to the regular Cherenkov light shining on all four anodes.  The number of anodes hit could provide an additional information to select out the window events. New 4-anode PMTs almost double the signal for the real events but produce significantly lower level signals for the window events (Fig.~\ref{fig2})~\cite{ref1,ref2,ref3}. 

\begin{figure}[htb]
\centering
\includegraphics[height=2in]{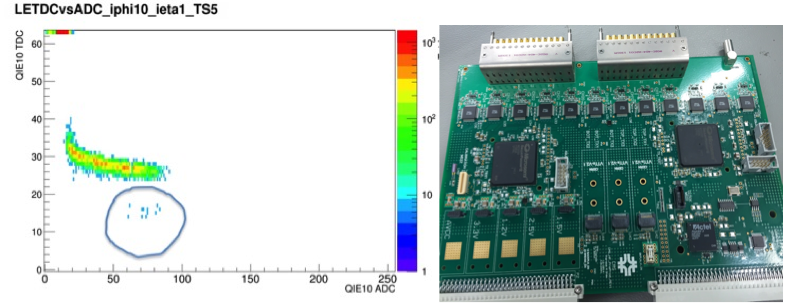}
\caption{ (a) Arrival time versus the pulse height the of the PMT signals (left) during the collisions at 13 TeV~\cite{ref4}. (b) New 2-channel readout HF-FE cards (right).}
\label{fig3}
\end{figure}

\begin{figure}[htb]
\centering
\includegraphics[height=2in]{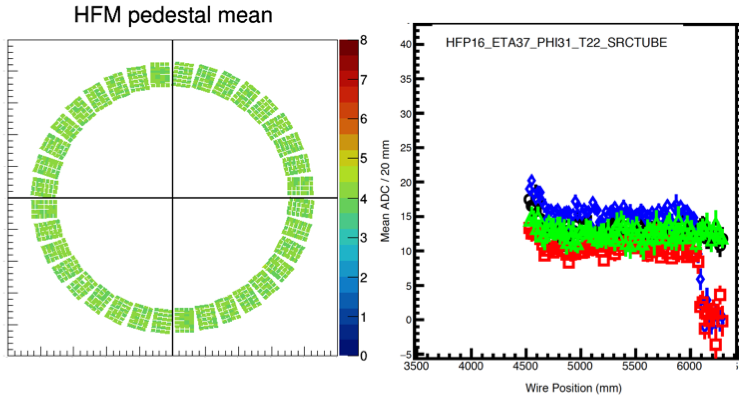}
\caption{Pedestal test results for HF- (left) and the $^{60}$Co source calibration scans observed in two channels for a tower in HF+.  Black and green points are for the EM channels, red and blue are for the hadronic channels (right)~\cite{ref5}.}
\label{fig4}
\end{figure}

Initially, the new PMTs are read out with the existing front-end electronic cards designed for single anode PMTs by combining the signals from all four anodes. New electronic cards were needed to utilize the additional channels. Due to cost considerations, the new HF Calorimeter Front-End Electronic cards are redesigned to read the four-anode PMTs in two channels. The new cards also have the new version of the ADC chip (QIE10 instead of QI8) to provide the timing information for the event arrival. Towards the end of the 2015 run period, such a redesigned HF-FE card was placed in the detector instead of an existing card. Pulse height and timing information provided by this new card confirmed that the window events produce large signals and come earlier (Fig.~\ref{fig3}a) (inside the circled area)~\cite{ref4}. A Turkish company in Istanbul (SIMPRO) assembled 200 of these cards (Fig.~\ref{fig3}b) last summer. They were tested and installed at the beginning of this year during the extended year-end technical stop (EYETS). While replacing all the front-end cards in the detectors, read-out boxes were refurbished and reconfigured to handle the additional information channels, both two-anode and timing signals. Improved calibration and control modules were also installed in the detectors. All the PMTs in the detectors were tested and calibrated by LEDs and $^{60}$Co source (Fig.~\ref{fig4}). Both detectors have been commissioned and ready for the collisions starting in May of this year (2017)~\cite{ref5}.

\begin{figure}[htb]
\centering
\includegraphics[height=1.8in]{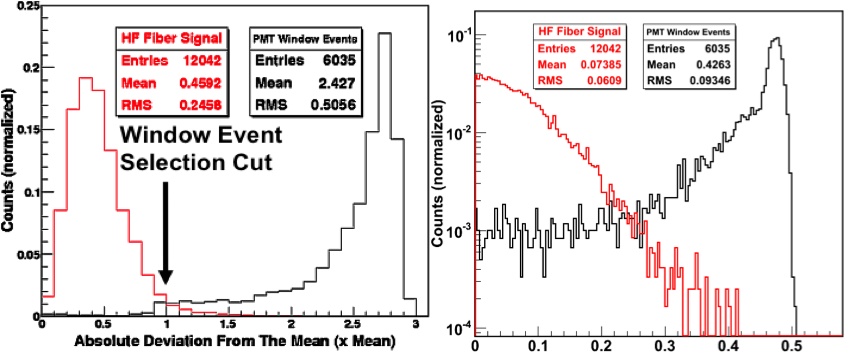}
\caption{ (a) Distributions of the difference between individual anode and the total signal normalized to the total signal (b) Same type of distributions for two channel read-out~\cite{ref1,ref2,ref6}.}
\label{fig5}
\end{figure}

\section{Software upgrade: algorithms to eliminate window events}

At higher luminosities, it will be difficult to eliminate the window events even with the new PMTs. Comparison of the individual anode signals may help eliminate the window events as shown in earlier test beam measurements at CERN. HF calorimeter is simulated by placing a bundle of the same fiber used in the HF detector behind a steel absorber, and the fibers are attached to the PMTs. Placing the PMTs in the muon beam simulates the window events. Histograms of the difference between the single anode signals and the mean of all four anodes are compared for both types of events. Distributions observed for window and real events are well separated (Fig.~\ref{fig5}a). Even with two channel readout these distributions still shows a good separation (Fig.~\ref{fig5}b))~\cite{ref1,ref2,ref6}. In two-channel readout, the signals from two anodes are combined and read out as a single signal. During the collision run starting in May 2017, algorithms will be developed and studied further. 


\end{document}